\documentstyle[twoside,fleqn,npb,epsfig]{article}
\def\asb{{\bar \alpha}_{\mbox{\scriptsize s}}}
\def\as{\alpha_{\mbox{\scriptsize s}}}

\def\nc{N_C}

\def\qt{q_t}

\newcommand{\AmS}{{\protect\the\textfont2
  A\kern-.1667em\lower.5ex\hbox{M}\kern-.125emS}}

\newcommand\hepph[1]{hep-ph/#1}

\newcommand\jhep[3]{{{\it J. High Energy Phys. }{\bf #1} (#2) #3}}

\newcommand\npb[3]{{{\it Nucl. Phys. }{\bf B #1} (#2) #3}}

\newcommand\plb[3]{{{\it Phys. Lett. }{\bf B #1} (#2) #3}}

\newcommand\prd[3]{{{\it Phys. Rev. }{\bf D #1} (#2) #3}}

\newcommand\prl[3]{{{\it Phys. Rev. Lett. }{\bf #1} (#2) #3}}

\title{
\begin{flushright}\normalsize
  \vspace{-1cm}
  Bicocca--FT--99--12\\
  hep-ph/9905308
  \vspace{0.5cm}
\end{flushright}
BFKL and CCFM final states%
\thanks{Talk presented at 7th International
  Workshop on Deep Inelastic Scattering, Zeuthen, Germany, April
  1999.}
}

\author{Gavin P. Salam\address{INFN --- Sezione di Milano, Via Celoria 
    16, Milano 20133, Italy}%
  \thanks{This work was supported by the EU Fourth Framework
    Programme `Training and Mobility of Researchers', Network `Quantum
    Chromodynamics and the Deep Structure of Elementary Particles',
    contract FMRX-CT98-0194 (DG 12-MIHT).}}

\begin{document}
\thispagestyle{empty}

\begin{abstract}
  I give a brief presentation of recent results on the equivalence
  of BFKL and CCFM small-$x$ final states, and discuss their
  implications for phenomenology.
\end{abstract}

\maketitle

\section{Introduction}

There are two basic approaches to the study of final states at
small-$x$. The BFKL equation \cite{BFKL} is derived in the
`Multi-Regge' limit,
i.e.\ %
assuming large rapidity intervals between successive emissions. This
guarantees the leading logarithms $(\as\ln x)^n$ (LL) for sufficiently
inclusive quantities (such as the total cross section, forward-jet
rates) but not necessarily for more exclusive final properties, such
as multiplicities or correlations.

A different approach, the CCFM equation \cite{CCFMa}, goes
beyond the `Multi-Regge' approximation, and explicitly treats
coherence (angular ordering) and soft emissions (the $1/(1-z)$ part of
the splitting function, where $z$ is the fraction of energy
remaining after a parton splitting). This guarantees the leading
logarithms for any small-$x$ observable, regardless of how exclusive
it is.

The main practical disadvantage of the CCFM equation compared to the
BFKL equation is that it is much more difficult to solve, both
numerically and analytically. It is therefore of some importance to
understand precisely in which situations the BFKL equation will give
the correct answer.

A few years ago it was shown by Marchesini that for quantities such as
multiplicities, the two equations differed at the level of double
logarithms (DL) of $x$, $(\as \ln^2 x)^n$ \cite{Pino95}. More recently
Forshaw and Sabio Vera \cite{FS} introduced a resolvability parameter
$\mu_R$ and showed (at fixed order, subsequently extended to all
orders by Webber \cite{Webber}) that $n$-resolved-particle ($n$-jet)
rates are the same in BFKL and CCFM at
leading DL order %
(all terms $(\as \ln^2 x)^m (\as\ln x \ln Q)^n$). Since multiplicities
are just a weighted sum of $n$-particle rates, but with $\mu_R=0$, one
is led to ask how these two, apparently contradictory, results are
related.

A second issue is that the above results were obtained without any
consideration of the soft emissions. What effect do they have? 

These questions were discussed in \cite{Salam} and the basic ideas are
presented in the next section.  Essentially, the inclusion of soft
emissions leads to all BFKL and CCFM predictions being identical at LL
level.  The phenomenological implications of this result are discussed
in sections 3 and 4.

\section{Theoretical properties of final states}

The fundamental property used in \cite{Salam} for the study of
final-state properties, is that in both the BFKL and the CCFM equations
it is possible to separate emissions which change the exchanged
transverse momentum $k$ (i.e.\ %
have the largest transverse momentum of all emissions so far ---
$k$-changing emissions) from those which don't ($k$-conserving). The
$k$-changing emissions are responsible for determining the cross
section, and can be shown, quite simply, to have the same structure in
BFKL and CCFM.

\begin{figure}[t]
  \begin{center}\vspace{-0.3cm}
    \input{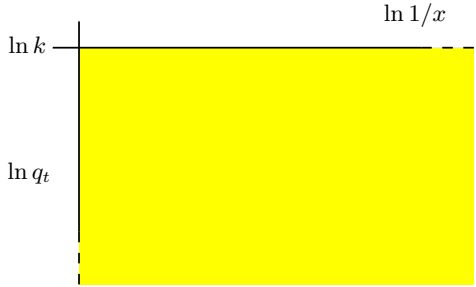}\vspace{-0.7cm}
    \caption{Distribution of BFKL emissions\vspace{-1.0cm}}
    \label{fig:bfkl}
  \end{center}
\end{figure}

It is the $k$-conserving emissions which are organised differently
between BFKL and CCFM, so I will consider just those.  For BFKL,
fig.~\ref{fig:bfkl} shows as the shaded region the distribution of
$k$-conserving emissions in $x$ (longitudinal momentum fraction of the
emitted gluon) and $\qt$ (transverse momentum of the gluon) space.
Emissions are independent with mean density
\begin{equation}
\label{eq:bfkldens}
\left \langle \frac{dn}{d\ln \qt \, d \ln 1/x} \right\rangle = 2\asb\,,
\end{equation}
where $\asb = \as\nc/\pi$.

\begin{figure}[t]
  \begin{center}\vspace{-0.3cm}
    \input{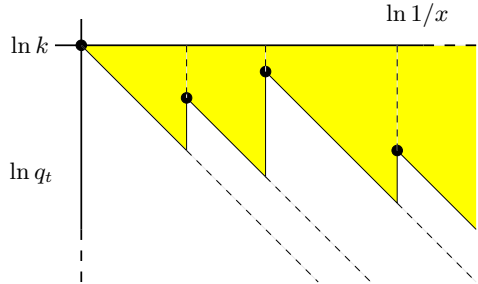}\vspace{-0.7cm}
    \caption{Distribution of CCFM emissions: discs are hard emissions, the
      shaded areas are the regions accessible to subsequent hard
      emissions. Soft emissions fill up the white
      space.\vspace{-1.0cm}}
    \label{fig:ccfm}
  \end{center}
\end{figure}

The CCFM equation has the extra constraint that emissions be ordered
in rapidity. This is illustrated in fig.~\ref{fig:ccfm} --- the
diagonal lines are constant rapidity ($\ln \qt/x$), and the
requirement of a given emission having a larger rapidity than the
previous one (being to the right of the diagonal line from the
previous one) eliminates the small-$\qt$ emissions. Mathematically
this translates into a mean density of emissions of 
\begin{equation}
  \label{eq:ccfmdens}
  \left \langle \frac{dn}{d\ln \qt \, d \ln 1/x} \right\rangle \simeq
  2\asb e^{-\asb \ln^2 \qt/k}\,,
\end{equation}
which differs from the BFKL result by a subleading factor (containing
no logarithms of $x$). So for finite $\qt$ one obtains the result that
BFKL and CCFM emission rates are the same at LL, in accord with the
results of Forshaw and Sabio Vera, and Webber \cite{FS,Webber}. But
integrating over all $\qt$ and $x$ to get the total multiplicity gives
\begin{equation}
  \label{eq:mult}
  \langle n\rangle \simeq \sqrt{\pi \asb \ln^2 x}\,,
\end{equation}
which is a double logarithm of $x$ just as found by Marchesini in
\cite{Pino95}. The message is that formally subleading 
transverse DLs $\as \ln^2 \qt$ play a fundamental role and can
be thrown away only in specific circumstances.

So far in the CCFM equation we have considered results including just
hard emissions, those from the $1/z$ part of the splitting function.
The inclusion of the soft emissions, those from the $1/(1-z)$ part of
the splitting function, changes the results radically, filling in the
regions between the dashed lines of fig.~\ref{fig:ccfm} in such a way
that the combination of soft and hard emissions turns out to
correspond to independent emissions with the density given by
(\ref{eq:bfkldens}), identical to the result from the BFKL equation.
This equivalence between BFKL and CCFM predictions holds at LL
accuracy (all terms $(\as\ln x)^n$) \cite{Salam}. There are actually
still some differences at subleading transverse DL %
level $\as\ln^2 \qt$ but they are confined to the end of the chain of
gluons and do not resum in such a way as to affect the LL results.

\section{Implications for phenomenology}
\label{sec:phen}
The above result on the LL equivalence of BFKL and CCFM final states
is a formal statement. It has relevance for analytical calculations of
the LL properties of final states, e.g.\ \cite{EW}.  But the BFKL and
CCFM equations have fundamentally different physical origins, and this
is reflected in differences at subleading order: the BFKL equation
(formulated as an evolution in $x$, for DIS) has a value of $z$
(energy fraction remaining after a parton splitting) which is
determined essentially by an arbitrary collinear cutoff $\mu$, present
for regularisation purposes:
\begin{equation}
  \ln \frac1z \sim \frac{1}{2\asb \ln k/\mu}\,.
\end{equation}
In the formal limit of small $\asb$, typical $z$ values are small and
there are no problems. But with $\mu/k \to0$, $\langle z \rangle$
becomes arbitrarily close to $1$. Since rapidities go as $\ln
z\qt/(1-z)$, the rapidities of the emitted gluons (even the hardest,
i.e.\ jets) depend, at subleading level, on the collinear
cutoff.\footnote{If one instead formulates the BFKL equation as an
  evolution in rapidity, as done for Monte Carlo implementations for
  the Tevatron \cite{MCBFKL} the problem disappears, but other serious
  difficulties arise for DIS, such as structure functions containing
  spurious transverse DLs.} In contrast, because of the explicit
treatment of coherence and the separation of soft emissions, the CCFM
equation never shows such pathological behaviour ($z$ is
well-defined), and so is a much better candidate for detailed
phenomenology, for example in the form of a Monte Carlo program such
as \textsc{smallx} \cite{SMALLX} (for an application to HERA data see
\cite{Jung}).

But the CCFM equation is not entirely free of problems: there are
subleading ambiguities in its implementation which can have large
effects on its predictions \cite{BMSS}. Additionally the CCFM equation
lacks an important symmetry: it works well evolving from a low
transverse scale to a high one (DIS), but not in the opposite
direction.  The symmetry issue is actually resolved in the Linked
Dipole Chain approach (LDC) \cite{LDC}, which like the CCFM equation
has a separation of hard and soft emissions. But the LDC does not
reproduce the BFKL cross section at LL level. It is not currently
clear whether this might be related to its problems in describing the
data.

\section{Outlook}

The present phenomenological situation is that none of the approaches
contains all of the physics that might be considered mandatory. Though
the formal LL equivalence of BFKL and CCFM final states is of limited
immediate relevance for phenomenology, it is an important step in our
general understanding of small-$x$ final states: together with
information from the NLL corrections \cite{NLL,NLLUnderstand} it gives
us a picture of the features required in future phenomenological
approaches.

\section*{Acknowledgements}
I am grateful to Yu. Dokshitzer, M. Ciafaloni, G. Marchesini and
S. Munier for discussions.

\end{document}